\documentclass[aps,prl,superscriptaddress,unsortedaddress,showpacs]{revtex4}

\usepackage{amsfonts}

\usepackage[pdftex]{graphicx}
\usepackage{epstopdf}
\usepackage{amsmath}
\usepackage{amssymb}
\usepackage{tikz}
\usepackage[page,titletoc]{appendix}
\usepackage[final]{pdfpages}

\definecolor{linkcolor}{rgb}{0,0,0.6} 
\usepackage[pdftex,colorlinks=true, pdfstartview=FitV, linkcolor= linkcolor, citecolor= linkcolor, urlcolor= linkcolor, hyperindex=true,hyperfigures=true]{hyperref} 

\begin{document}  

\title{Fast equilibrium switch of a micro mechanical oscillator}

\author{Anne Le Cunuder}
\affiliation{Université de Lyon, CNRS, Laboratoire de Physique de l'\'Ecole Normale Sup\'erieure, UMR5672, 46 All\'ee d'Italie, 69364 Lyon, France.}

\author{Ignacio A. Mart\'inez }
\affiliation{Université de Lyon, CNRS, Laboratoire de Physique de l'\'Ecole Normale Sup\'erieure, UMR5672, 46 All\'ee d'Italie, 69364 Lyon, France.}

\author{Artyom Petrosyan}
\affiliation{Université de Lyon, CNRS, Laboratoire de Physique de l'\'Ecole Normale Sup\'erieure, UMR5672, 46 All\'ee d'Italie, 69364 Lyon, France.}

\author{David Gu\'ery-Odelin}
\affiliation{Laboratoire de Collisions Agr\'egats R\'eactivit\'e, CNRS UMR 5589, IRSAMC}

\author{Emmanuel Trizac}
\affiliation{LPTMS, CNRS, Univ. Paris-Sud, Universit\'e Paris-Saclay, 91405 Orsay,
  France}

\author{Sergio Ciliberto}
\affiliation{Université de Lyon, CNRS, Laboratoire de Physique de l'\'Ecole Normale Sup\'erieure, UMR5672, 46 All\'ee d'Italie, 69364 Lyon, France.}
\email{sergio.ciliberto@ens-lyon.fr}

\begin{abstract}
We demonstrate an accurate method to control the motion of a micromechanical oscillator in contact with a thermal bath. The experiment is carried out on the cantilever tip of an Atomic Force Microscope (AFM). Applying an appropriate time dependent external force, we decrease the time necessary to reach equilibrium by two orders of magnitude compared to the intrinsic equilibration time. Finally, we analyze the  energetic cost of such a fast equilibration, by measuring with $k_BT$ accuracy the energy exchanges along the process. 
\end{abstract}

\maketitle

\textit{Introduction}. 
The last decade witnessed spectacular advances in the fabrication and control of high-quality micromechanical oscillators. They are nowadays widely used in applications including timing, synchronization, high precision sensing of force, acceleration and mass. They even provide an interesting connection between quantum resources dedicated to quantum state manipulations and resources for transmitting quantum states \cite{pep1,pep2,Holland}.
 
Most applications involve micromechanical oscillators in the underdamped regime and in contact with a thermal bath.  In the present letter, we implement a generic method to speed up the transition between two equilibrium states of such a micromechanical oscillator in the limit where the relevant description is that provided by the 1D underdamped harmonic oscillator in the presence of thermal noise. 
Such a system evolves towards a new equilibrium state by dissipating energy along an oscillating dynamics whose amplitude decreases with a relaxation time of 
$\tau=m/\gamma$, where $m$ is the oscillator mass and $\gamma$ the viscous coefficient, which depends on the  surrounding medium and the probe geometry. 
The reduction of the duration time of the transient regime to an arbitrary time $t_f\ll \tau$ is an important issue for applications, 
such as for example Atomic Force Microscopy (AFM), which has become a pivotal tool in the experimental study of biological systems,  material science, polymer physics  etc.
Many  experiments are done in gaseous media, which  increases the quality factor and produces long transients. An arbitrary acceleration of  the equilibration time of AFM cantilevers is the basis 
of the high speed AFM, and it has been achieved in particular
using feedback techniques \cite{humphris2005,Nony,Devasia} or changing the viscoelastic behavior of the cantilevers \citep{adams2015}. 
Alternatively, it has been recently  shown, in the case of an overdamped system, that fast relaxation can be obtained by using an appropriate driving force which remains efficient 
even in a 
very noisy environment. This result was obtained on a Brownian particle trapped by  optical tweezers and the new equilibrium was reached 100 times faster than the natural 
equilibration time \cite{martinez2015faster}. Here, we  generalize this  idea, referred to as Engineered Swift Equilibration (ESE), to underdamped systems, using as micro 
mechanical oscillator the cantilever tip of an atomic force microscope. 
We also measure directly the energy needed in the course of the transformation to accelerate the process. 
Our approach is therefore of feed-forward type, and in that, belongs to a category
of techniques known in the engineering community as {\em input shaping} \cite{Singer,Devasia}.

Specifically, we propose  an  ESE protocol, which does not require any feedback and which is based only  on a statistical analysis of the cantilever tip  position $x(t)$, 
whose dynamics is described with a rather good accuracy by a second order  Langevin equation:
\begin{equation}
m \ddot{x}= - \gamma \dot{x} -\kappa x +{F(t)}+ \zeta(t)
\label{eq:langevin}
\end{equation}
where $\kappa$ is the stiffness of the system and $F$ the external applied force.  $\zeta$ is a white noise delta correlated in time: 
$\langle\zeta(t)\zeta(t')\rangle=2\gamma k_B T \delta(t-t')$. The resonant frequency $\omega_o=\sqrt{\kappa/m}$ is the frequency of the first cantilever mode. 
The process that we want to speed up is the transition of the cantilever tip from an initial  equilibrium position $x_i$  to a new one $x_f$, obtained by applying a 
time dependent force $F(t)$.

In the case of an underdamped oscillator in the presence of  thermal fluctuations, the equilibrium velocity and position probability distribution function (pdf) $\rho_{eq}(x,v,t)$ 
reads as $\rho_{eq}(x,v) =\frac{1}{Z} \exp\left[-\frac{\kappa x^2}{2k_BT}-\frac{Fx}{k_BT}\right] \exp\left[-\frac{mv^2 }{2k_BT}\right]$, where $\kappa$, $m$ and $T$ are fixed all 
along the protocol and $Z$ is the partition function.
Once a parameter is changed, for example the external force $F$, the Kramers equation  (see supplementary information \cite{SuppInf} ) gives us the evolution of the pdf. 
By tuning appropriately the strength of $F$ as a function of time, it is possible to force the system to equilibrate in a given  time $t_f$. This is the spirit of the ESE protocols. 

Specifically, in our  ESE process  (see supplementary information for an accurate derivation\cite{SuppInf} ) the force evolves according to a polynomial equation in the normalized time $s=t/t_f$ as
\begin{eqnarray}
{{F}(s) \over \kappa  \  x_f}&=& s^3 (10-15s+6s^2)+ \notag \\ &\ +& \frac{\gamma}{\kappa t_f}(30 s^2-60s^3+30s^4)+\notag   \\	&\ +&\frac{m}{\kappa t_f^2}(60s-180s^2+120s^3)
\label{eq:Force}
\end{eqnarray}
with boundary conditions $F(t)=0$ for $t<0$ and $F(t)/\kappa= F_f/\kappa=x_f$ for $t>t_f$. Interestingly, the protocol dependence on the final position $x_f$ is separable, 
and keeps its shape under different traveling distances. The protocol contains three terms with different dependence on the intrinsic parameters of the system $\kappa$, $\gamma$, $m$  whose importance is pondered under different powers of $t_f$. As demonstrated in the following, the system will reach the desired equilibrium state from its initial equilibrium state using this protocol and in the time interval $t_f$ that we choose. 

\begin{figure}[!ht]
\centering
\includegraphics[width=0.5\columnwidth]{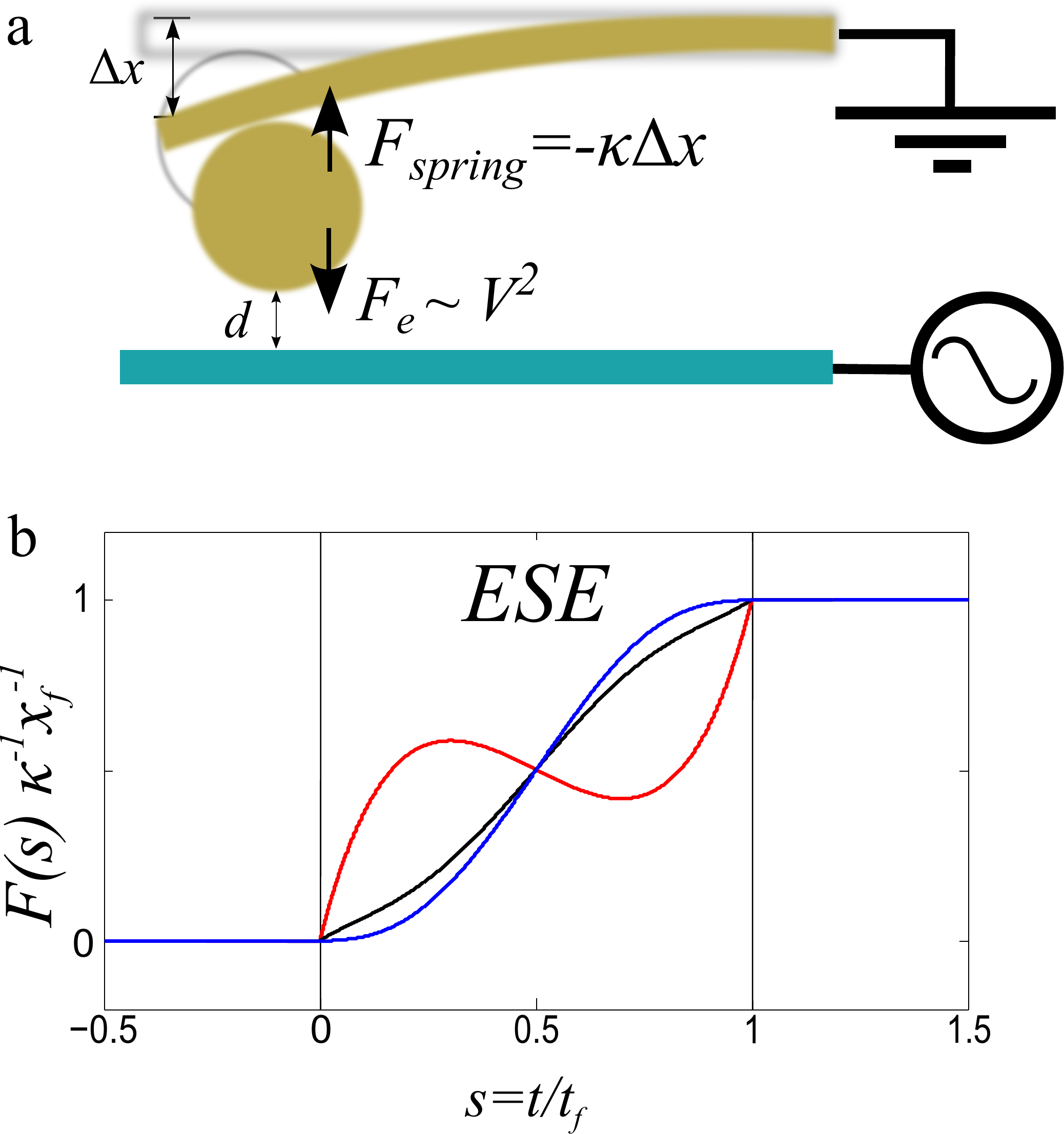} 
\caption{ \textbf{a}- Sketch of the experimental setup. The cantilever-sphere system  is connected to ground, while the surface is connected to the signal generator. The external force is applied by the voltage difference between them. 
\textbf{b}- Force ESE protocol for different final  times $t_f$=0.2 ms (red), 0.5 ms (black) and 2 ms (blue) as a  function of the normalized time $s=t/t_f$. If we reduce enough the protocol time, the inertial term of the protocol becomes dominant and yields a non monotonous force, see the red curve.}
\label{fig:fig1}
\end{figure}

\textit{Experimental setup}.
The sketch of the experimental setup is illustrated in Fig.\ref{fig:fig1}a). The oscillator under consideration is a silicon cantilever (size 500 $\mu$m $\times$ 30$\mu$m $\times$ 2.7$\mu$m, 
NanoAndMore) with a polystyrene sphere (Sigma-Aldrich, $R=75\mu$m) glued on its tip. We will refer to this ensemble sphere-cantilever as the probe.  
The whole probe and the flat surface, facing the sphere,  are coated with a 100nm thick gold layer.  The experiment is done in nitrogen atmosphere at room temperature $T=$ 300 K and pressure $p=$1 bar. 
Therefore, the viscosity is very low, which gives a completely underdamped dynamics. The surface-sphere distance  $d$ can be tuned  using an electronically controlled  piezostage (Piezo Jena).  The position $x(t)$  of the cantilever tip is measured by  a highly sensitive interferometer with subpicometer resolution and high speed acquisition $f_{\rm acq}=200$ kHz \cite{paolino2013}. The stiffness, viscosity and mass are intrinsic parameters of the oscillator and are calibrated using the Brownian motion of the thermally excited probe, in this specific case $\kappa=(2.50\pm 0.50){\rm N/m}$,  $\gamma=(1.00\pm0.30) \cdot10^{-6}$ Ns/m and $m=(8.37\pm 0.16)\cdot10^{-9}$ kg. Hence, the resonance of the first mode of the probe is $\omega_o = 17.3$ krad/s.

An attractive electrostatic interaction is generated by applying a voltage difference $V(t)$ between the surface and the sphere. The sphere-surface force  can be written as $F=4\pi\epsilon_0 R V^2/d$ for $d\ll R$ \cite{capacitance}, where $\epsilon_0$ is the dielectric constant, $d$  the distance between the 
sphere and the surface and $R$ the radius of the bead. Therefore, we can write $F=\Lambda V^2$, where $\Lambda=(1.71\pm 0.01)\cdot 10^{-10}{\rm N/V^2}$ 
is the calibration factor obtained from the equilibrium relation $\kappa \Delta x=\Lambda V^2$, where $\Delta x$ is the displacement of the cantilever 
once we apply a voltage $V$ \cite{gomez2010}. In practice, the voltage $V(t)$ is produced  by an arbitrary signal generator (Agilent 33522) at 2 
MHz sampling rate. Experiments were performed  at a distance $d> 1\mu$m  and for 
a maximum required displacement below $\Delta x=|x_f-x_i| < 3$nm. The  condition $\Delta x \ll d$ {allows us to neglect the dependence of $\Lambda$ on $x$, 
which opens the way to a simple implementation of the ESE protocol,  with an $x$ independent  force, see Eq.~(\ref{eq:Force}).}

\textit{Experimental protocols}.
The time dependent  behavior of $F(t)/(\kappa x_f)$ needed to equilibrate the probe in a time $t_f$ is obtained by inserting in Eq.~\ref{eq:Force} the experimental values of the 
parameters. 
The computed time evolutions of $F(t)$ are plotted in Fig.\ref{fig:fig1}b),  for various $t_f$. {Decreasing $t_f$ below some threshold  (for our experiment,
this threshold is 0.23\, ms), the behavior is no longer monotonous }.  \\
In order to emphasize the main features  of  ESE, we compare it to  a standard step  protocol (STEP)  in which  we instantaneously change $F(t)$ from $F_i=0$ to the final value $F_f=\kappa \, x_f$.
In the absence of noise, the response of the system to STEP forcing obeys the equation $x(t)/x_f=1-\exp(-\xi\omega_o t)\sin(\sqrt{1-\xi^2}\omega_o t+\phi)/\sqrt{1-\xi^2}$, where $\xi=\gamma/(2\sqrt{m\kappa})$ 
and $\phi=\cos^{-1}\xi$.  From this time evolution, one can fix a reference  velocity  $v_0=\frac{x_f}{\sqrt{1-\xi^2}}\xi\omega_o=(20.9 \pm 0.2){\rm nm/ms} $, 
which we use to compare quantitatively STEP and ESE responses.

Examples of the time evolution of $x(t)$ and $\dot x(t)$ for the two protocols  are plotted in Fig. \ref{fig:fig2} when the equilibrium position is changed from $x_i=0$ to $x_f=0.5$ nm. In this specific illustration, 
we choose for ESE $t_f=2\,$ms for which  the needed  $F(t)$ is  plotted  in  Fig.~\ref{fig:fig1}b) (blue line).  In Fig.~\ref{fig:fig2}, we compare the  
STEP and  ESE protocols by plotting for each of them, a single realization (blue lines) and the mean  response (red lines) obtained by averaging over 5000 realizations of the protocols. 
Within the  STEP protocol (Figs.~\ref{fig:fig2} b) and d)), both $x(t)$ and $\dot x(t)$  do not relax up to more than several $\tau$. This has  to be compared to ESE (Figs.~\ref{fig:fig2} a) and c)), for which the system reaches the target position $x_f$ in the desired timelapse ($t_f=2$ ms);  
this is about two order of magnitude faster than STEP. Note that the velocity scale for the velocity along ESE is five time expanded with respect to STEP. Remarkably, the ESE turns out to be very efficient even  at the level of a single realization.

{However, the ESE formulation put to work here cannot be operational for too small values of $t_f$.
This can be observed by comparing results with $t_f$ ranging from 0.2 ms to 10 ms.} As shown in Fig.~\ref{fig:fig3}, where the ensemble averages of 5000 trajectories with $t_f=(0.2,0.5,2.0)\,$ms 
are presented, the response of the systems to ESE protocol is excellent  as long as $t_f>t_{\rm osc}=2\pi/\omega_o\simeq0.4{\rm ms}$. 
{For times shorter than $t_{\rm osc}$, the ESE response 
remains signficantly superior to its STEP counterpart,} but it  begins to deteriorate with the occurrence of small damped oscillations.  The reason of these residual oscillations  lies 
in  the modeling of the probe dynamics. Indeed, Eq.~\ref{eq:langevin}  describes well the probe dynamics only for frequencies smaller than the first longitudinal mode frequency  of 
the micro mechanical oscillator. When  $t_f<t_{\rm osc}$, the high order modes are excited too and Eq.~\ref{eq:langevin} does not describe properly the dynamics of the tip.  The effect 
is visible in the proper trajectory or velocity, see Fig. \ref{fig:fig3} for $t_f=0.2{\rm ms}$.
However, in spite of this small residual error,  the improvement for equilibration speed of ESE  is still quite appreciable in this regime,  as it can be easily checked by comparing  
the STEP response in \ref{fig:fig2}b)   with \ref{fig:fig3}a). 
{We emphasize that the feature addressed here is not a deficiency of the ESE method at such,
but a consequence of its implementation on an equation that becomes inacurrate at high frequency. 
This is the intrinsic limitation of ESE which works as a far as the mathematical model of the physical system is accurate enough

\begin{figure}[!ht]
\centering
\includegraphics[width=0.5\columnwidth]{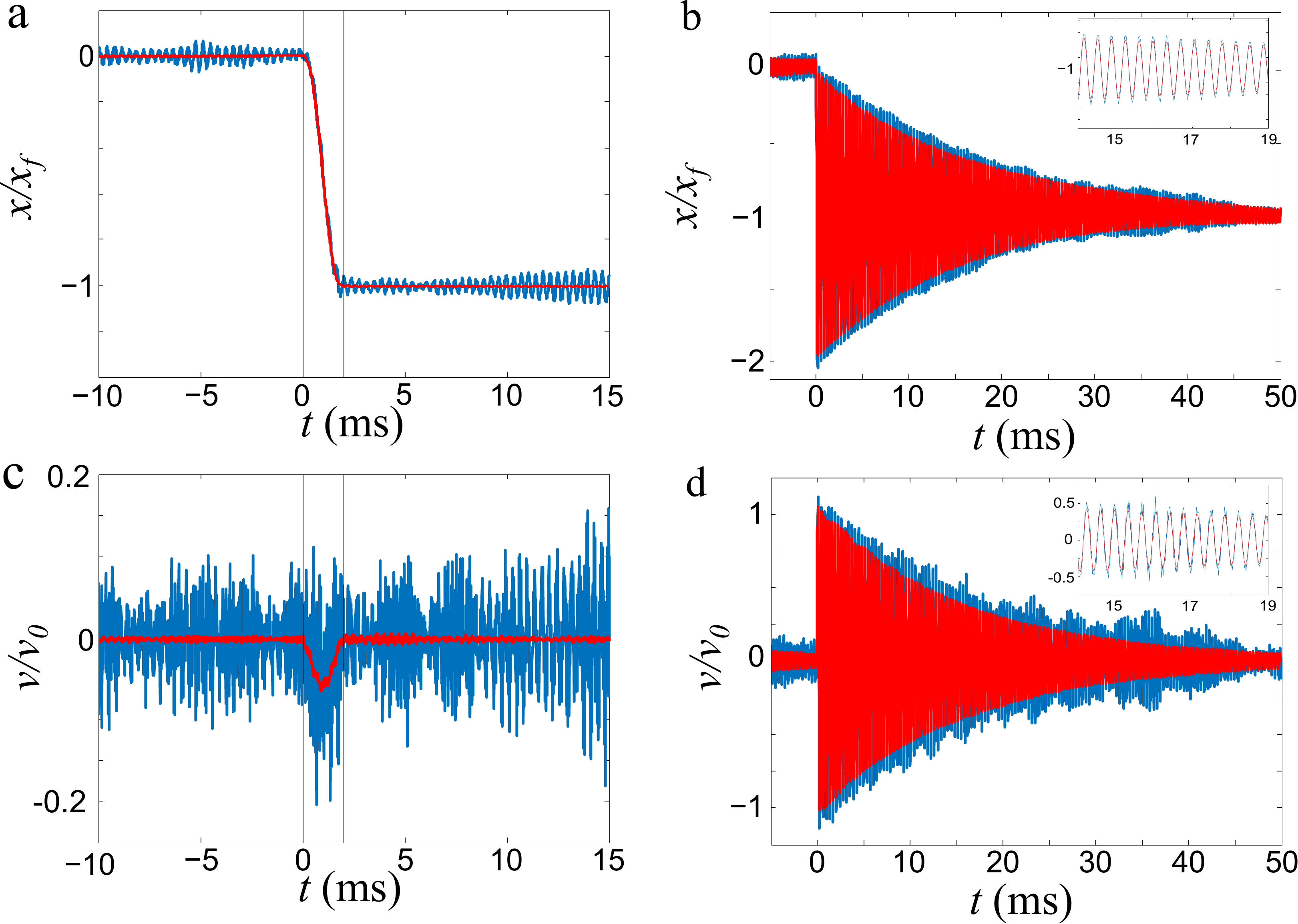} 
\caption{\textbf{Dynamics of the system along the STEP and ESE protocols. All processes start at $t=0$ ms.} 
\textbf{a}- Position evolution along the ESE process with $t_f=2.0$ ${\rm ms}$.
\textbf{b}- Position evolution along the STEP protocol. Intrinsic oscillations of the cantilever are much faster than the dissipation process what makes difficult to distinguish the trajectory. The inset provides a magnification of a small region.  
\textbf{c}-  Normalized velocity $\dot x(t)/v_o$  as a function of time, along the ESE process.
\textbf{d}- Normalized velocity velocity evolution along the STEP process. The inset increases the time resolution, to observe the intrinsic oscillations.  
All figures show the dynamics of a single realization (blue) and the ensemble average over 5000 realizations (red). Vertical black solid lines in \textbf{a} and \textbf{c} 
represent the limits of the ESE protocol
 ($t_f=2\,$ms).}
\label{fig:fig2}
\end{figure}

\begin{figure}[!ht]
\centering
\includegraphics[width=0.5\columnwidth]{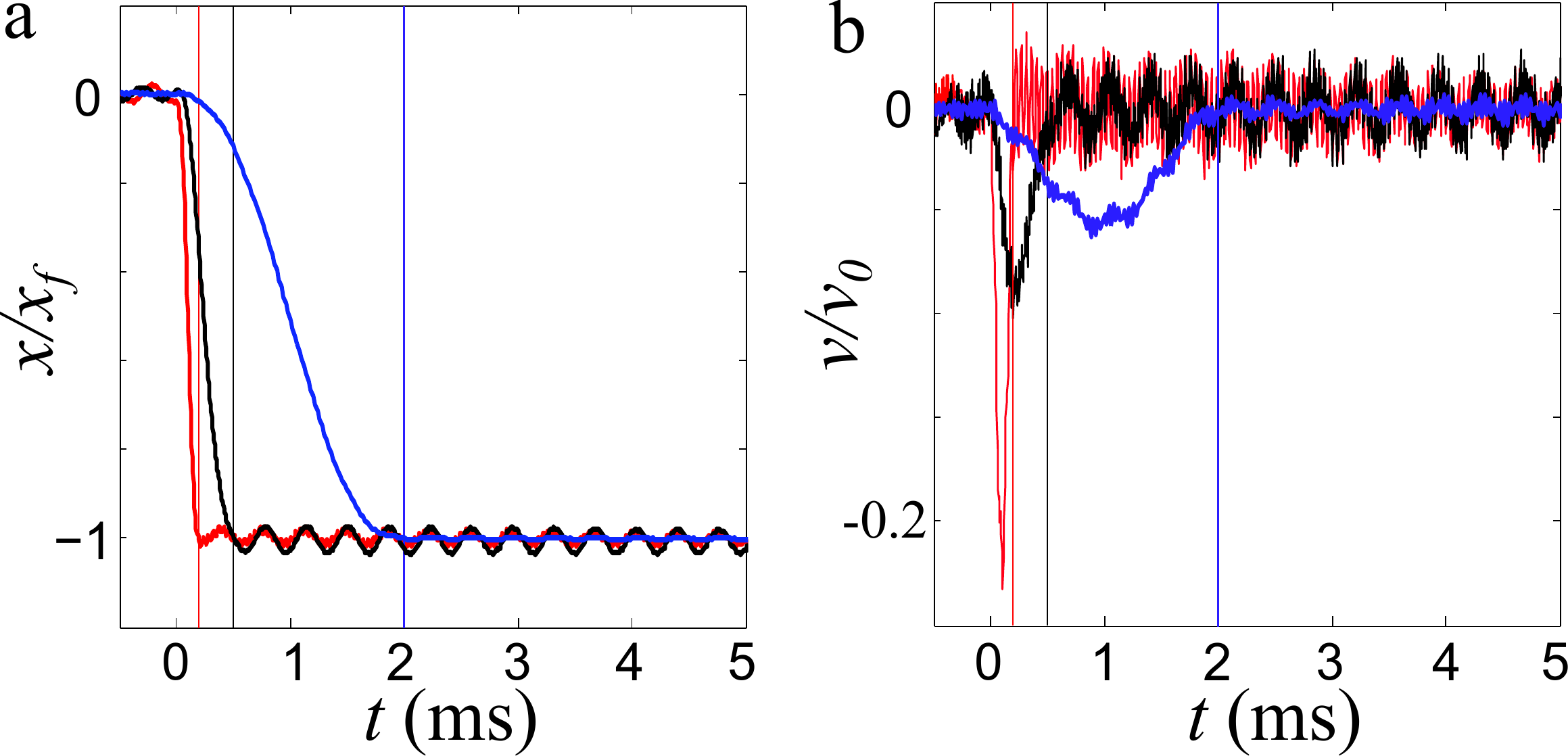} 
\caption{\textbf{Comparison of the  ESE protocols at various $t_f$}. 
\textbf{a}- Ensemble average of the trajectory over 5000 realizations for $t_f=0.2 {\rm ms}$ (red), $t_f=0.5 {\rm ms}$ (black) and $t_f=2 {\rm ms}$ (blue). 
The higher modes starts to dominates the dynamics of the system once the protocol is shorter than $2\pi/\omega_o\simeq 0.4\,$ms. 
\textbf{b}- Ensemble average of the normalized velocity for the same times than a). The normalized velocity $\dot x(t)/v_o$  allows a better vision of the higher orders effect. 
In both figures, the vertical lines represent the end of each protocol.  All processes start at $t=0$ ms.
}
\label{fig:fig3}
\end{figure}

\textit{Energetic study}.
A relevant question deals with the measurement of the energy dissipated for various values of $t_f$. Indeed a good characterization of the energetics of small systems is essential 
to understand their  time  evolution, their limits and their  interactions with the environment. Even if the energy exchange is comparable with the intrinsic thermal noise, this heat 
release is important in small devices, either  natural,  such as enzimes \cite{riedel2015heat},  or artificial,  such as thermal nanoengines \cite{martinez2016}.  
Our system has a total energy $E =U +K = \frac{1}{2} \kappa x^2 - Fx + \frac{1}{2}m v^2$, where $U$ and $K$ correspond to the potential and the kinetic energy, respectively. 
The stochastic energy received by the system along a single trajectory can be expressed as: 
 \begin{equation}
 \Delta E = \int_0^{t_f} \left[ \frac{\partial E}{\partial F} \dot{F} + \frac{\partial E}{\partial x} \dot{x} + \frac{\partial E}{\partial v} \dot{v} \right] dt~.
 \label{eq:ene}
 \end{equation}
Following Sekimoto\cite{seki2010,SuppInf}, we identify the first term in the rhs of Eq.~\ref{eq:ene} with the stochastic work $\delta W$. 
The heat $\delta Q=\delta E-\delta W$ splits into two contributions $\delta Q_x$ and $\delta Q_v$ which correspond to potential and kinetic heat respectively \cite{seki2010}. 
The value of the dissipated heat at the end of the protocol has to correspond to the difference between the exerted work and the difference of free energy between the 
initial and final state $\Delta {\cal F} $. The free energy difference is $\Delta {\cal F}= \Delta U - T\Delta S_{eq}$. 
As the difference of the entropy of the system $\Delta S_{eq}$ between the initial and final state is zero {(position-wise, the statistical distribution
is simply shifted by a quantity $x_f=F_f/\kappa$) and since $\Delta E=\Delta U$ due to the isothermal condition, the free energy difference is 
$\Delta {\cal F}= \Delta U=F_f^2/2\kappa=W+Q$ \footnote{In all this discussion, $W$ and $Q$ refer to the {\em mean} work and heat.}.}

In Fig.~\ref{fig:fig4}, the evolution of the energetics is shown for two different ESE times, $t_f=0.2\,$ms and $t_f=2\,$ms 
\footnote{{In spite of the limitation of our ESE implementation for $t_f<t_{\rm osc}$, we estimate the energetics in this regime as well. 
Indeed, we checked that  the measured time evolutions of $W$ and $Q$  
coincide quite well with those of the numerical simulation of Eq.~\ref{eq:langevin} for the ESE protocols. This means that the residual errors remain small even when $t_f<t_{\rm osc}$, 
as we have shown in Fig.\ref{fig:fig3}}}. Due to the low viscosity of the environment, dissipation lies within the detection limit of our experiment, and no significant changes are provided in the final value of the work needed to execute the protocol. However, there is a significant difference in the evolution of the energetics when we work above or below of the resonant frequency. When we work at time shorter than the period of oscillation, the protocol becomes non monotonous to compensate for the inertial term in the dynamics (see Fig.\ref{fig:fig1}b). Therefore, the work is not growing continuously, but becomes negative for a small time interval in the course of the transformation. This means that  the system shall exercise work on the environment to ensure a relaxation on a very short amount of time.  In Fig.~\ref{fig:fig3}b),  the total heat is plotted as a function of time. There are  significant changes in the time evolution but not in the final value. The total heat is the sum of the potential and the kinetic one, shown in Fig. \ref{fig:fig3}c) and d) respectively. As the cantilever moves from an equilibrium position to another rapidly, a high speed is required 
and a large amount of heat is absorbed via the  kinetic energy. Figures \ref{fig:fig3}c) and d) also show how the heat is dissipated via the potential energy and not via kinetic energy (see the non zero value in the final times of Fig.~\ref{fig:fig3}c) and the zero value in Fig.~\ref{fig:fig3}d). All graphs correspond to ensemble averages over 5000 realizations.

\begin{figure}[!h]
\centering
\includegraphics[width=0.5\columnwidth]{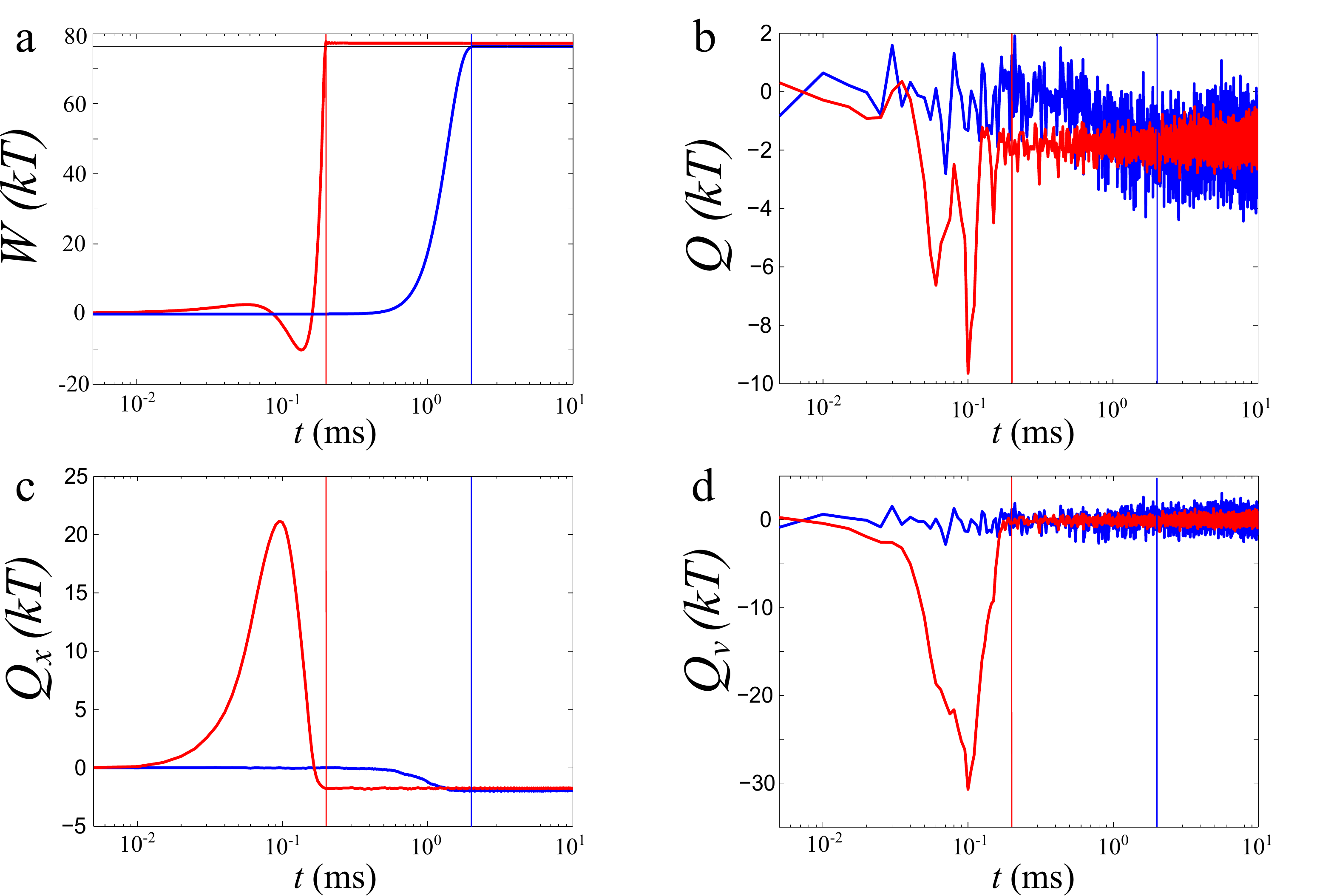}
\caption{ \textbf{Energetics of the system along the ESE route.}
\textbf{a} Average value of the cumulative work as a function of time for different protocol times $t_f$. Horizontal black line represents the difference of free energy between the initial and final state $\Delta \mathcal{F}$.
\textbf{b} Average value of the cumulative total heat.  
\textbf{c} Average value of the cumulative potential heat.
\textbf{d} Average value of the cumulative kinetic heat. Its importance reduces once we increase the protocol time. The kinetic heat is compensated with the potential one, showing the classical oscillation of a classical harmonic oscillator. In the four graphs, red lines are for $t_f$=0.2ms and blue lines for $t_f$=2 ms.}
\label{fig:fig4}
\end{figure}

\textit{Conclusions}.  
We have shown how high speed AFM could be designed using ESE protocols. A gain of two orders of magnitude in time has been demonstrated. 
The bound ultimately faced has to do with the limit of the modeling of the cantilever tip by a simple Langevin oscillator with a single resonant frequency. 
By reducing the operating time of AFM or Optical Trap, one enlarges the frequency window in which the system under scrutiny (biomolecule, material, enzyme, ...) can be probed. Our formalism can be readily applied to optical traps operating under vacuum and for which inertia become predominant \cite{Raizen}. 
We can also imagine protocols where the free parameter is the distance between the tip and the surface, modulated by a high accuracy piezoelectric device. Finally, the ESE protocols could be combined with standard feedback techniques to decrease further the operating time.

\bibliographystyle{ieeetr}
\bibliography{eseAFM}

\newpage
\newpage



\section*{Supplementary material}

# Fast equilibrium switch of a micro mechanical oscillator
# Supplementary Informations

Anne Le Cunuder,[1] Ignacio A. Martínez,[1] Artyom Petrosyan,[1]
David Guéry-Odelin,[2] Emmanuel Trizac,[3] and Sergio Ciliberto[1]

[1]Universit? de Lyon, CNRS, Laboratoire de Physique de l'École Normale Supérieure,
UMR5672, 46 Allée d'Italie, 69364 Lyon, France.
[2]Laboratoire de Collisions Agrégats Réactivité, CNRS UMR 5589, IRSAMC
[3]LPTMS, CNRS, Univ. Paris-Sud, Université Paris-Saclay, 91405 Orsay, France

PACS numbers:

## ENERGETICS MEASUREMENTS

For a system in contact with a heat bath described by a classical Langevin equation, stochastic energetics (Ref.14 of the main text) provides the framework to quantify work and dissipation. Along these lines, we calculate the thermodynamics quantities from measurements of the position of the sphere in the course of the transformation for a given protocol. The sphere attached to the cantilever has a total energy:

$$E = U(x,F) + K(v) = \frac{1}{2}\kappa x^2 - F\,x + \frac{1}{2}mv^2 \quad (1)$$

where $U(x,F)$ is the potential energy, $K(v)$ the kinetic energy and $F$ the electrostatic force applied to the system.

The elementary work done on the system reads

$$dW = \frac{\partial E}{\partial F} dF = -x dF, \quad (2)$$

and yields the following integrated expression over the duration of the protocol

$$W(t_f) = \int_0^{t_f} \frac{\partial E}{\partial F} \circ \frac{\partial F}{\partial t} dt \quad (3)$$

where ∘ denotes Stratonovich integral.

Similarly, the two contributions to the heat are given by:

$$dQ_x = \frac{\partial E}{\partial x} dx = \kappa x dx - F dx \quad (4)$$

and

$$dQ_v = \frac{\partial E}{\partial v} dv = mv dv. \quad (5)$$

We thus get, for the whole protocol duration

$$Q(t_f) = \int_0^{t_f} \frac{\partial E}{\partial x} \circ \frac{\partial x}{\partial t} dt + \int_0^{t_f} \frac{\partial E}{\partial v} \circ \frac{\partial v}{\partial t} dt. \quad (6)$$

It should be emphasized that the data reported in Fig.4 are the averages of these fluctuating quantities, taken over 5000 realizations of the experiment.

## ESE PROTOCOL

As a first approximation, the system made of the AFM cantilever plus the sphere can be modeled by a simple harmonic oscillator of resonant frequency $f_0 = (\kappa/m)^{1/2}/(2\pi) = 2750$ Hz. The validity of our model breaks down for frequencies exceeding $f_0$.

We monitor in real time the evolution of the system during the relaxation towards the new equilibrium state, and get therefore the expression for the probability density distribution $P(x,v,t)$. Its dynamics obeys the Kramers equation associated to the Langevin equation Eq.(1) in the main text :

$$m[\partial_t + v\partial_x - \omega_o^2(x-x_0)\partial_v]P = \gamma \partial_v[vP + k_B T \partial_v P] \quad (7)$$

where $\omega_o^2 = \kappa/m$ and $x_0(t) = F(t)/\kappa$. To work out the appropriate function $F(t)$, we proceed as follows: we find out a specific exact solution $P_e(x,v,t)$ of (7) that fulfills our boundary conditions $x_0(t_i) = 0$ and $x_0(t_f) = x_f$, and infer from this solution the external force that shall be applied.

For the sake of simplicity, we look for a solution of the form

$$P_e(x,v,t) = \exp\left[-\alpha(x,t) - \beta v^2 - \delta(t)v\right], \quad (8)$$

with $\beta = m/(2k_B T)$ constant. Combining Eqs. (7) and (8), we deduce that the function $\alpha(x,t)$ should be of the form $\alpha(x,t) = \alpha_0(t) + \beta\omega_o^2 x^2 + \Delta(t)x$ with $\Delta(t) = -\dot{\delta} - \gamma\delta/m - 2\beta\omega_o^2 x_0$. The term $\alpha_0(t)$ accounts for the normalization. Noticing that $\dot{\Delta} = \omega_o^2\delta$, we have a direct relation between $x_0(t)$ and $\Delta$:

$$x_0(t) = -\frac{1}{2\beta\omega_o^2}\left(\Delta + \frac{\gamma}{m\omega_o^2}\dot{\Delta} + \frac{1}{\omega_o^2}\ddot{\Delta}\right). \quad (9)$$

For consistency, we set the boundary conditions of the $\Delta$ function as: $\Delta(0) = 0$ and $\Delta(t_f) = -2\beta\omega_o^2 x_f$, $\dot{\Delta}(0) = \dot{\Delta}(t_f) = 0$ and $\ddot{\Delta}(0) = \ddot{\Delta}(t_f) = 0$. To fulfill those six constraints, we choose a polynomial of order five:

$$\Delta(t) = 2\beta\omega_o^2 x_f\, s^3\left(-10 + 15s - 6s^2\right), \quad (10)$$

with the dimensionless time $s = t/t_f$, varying from 0 to 1. From Eq. (9), we infer $x_0(t)$, and thus the expression $F(t)$ of Eq.(2) of the main text. Note that the

ESE protocol is worked out here in its simplest setting in which a harmonic potential has been used and the dynamics is controlled by only one parameter, i.e. $F(t)$ in this specific example. This is the reason why the results can be readily recovered directly from the Langevin equation, averaging over noise realizations to work with the mean position $\langle x \rangle$. Imposing the desired evolution for this quantity leads directly to Eq. (9), which can be supplemented with the polynomial choice (10) [1, 2].


\end{document}